\documentclass[journal=jacsat,manuscript=article]{achemso}
\usepackage{soul,xcolor}
\setkeys{acs}{doi=true}

\usepackage[normalem]{ulem}
\usepackage{chemformula} 
\usepackage[T1]{fontenc} 

\usepackage{braket}
\usepackage{lineno}

\usepackage[english]{babel}

\usepackage[pdftex,colorlinks=true,bookmarks=false,citecolor=blue,urlcolor=blue]{hyperref} 
\hypersetup{pdfstartview={FitH},pdfpagemode={UseNone}, breaklinks=true,
            colorlinks,linkcolor=blue, citecolor=blue, urlcolor=blue,
             bookmarksopen=true, pdfnewwindow=true}
\usepackage[all]{hypcap}
\newcommand{\doi}[1]{\href{https://doi.org/#1}{\nolinkurl{#1}}}

\graphicspath{{pict/}{figures/}{}}

\author{Jinyong Ma} 
\altaffiliation{equal contribution}
\email{jinyong.ma@anu.edu.au}
\author{Jihua Zhang} 
\altaffiliation{equal contribution}
\author{Yuxin Jiang} 
\author{Tongmiao Fan}
\author{Matthew Parry} 
\author{Dragomir~N.~Neshev} 
\author{Andrey A. Sukhorukov} 
\affiliation
{ARC Centre of Excellence for Transformative Meta-Optical Systems (TMOS),\\ Department of Electronic Materials Engineering, Research School of Physics,\\ The~Australian National University, Canberra, ACT 2601, Australia}

\title{Polarization engineering of entangled photons from a lithium niobate nonlinear metasurface}

\begin{document}
\sloppy

\begin{abstract}
Complex polarization states of photon pairs are indispensable in various quantum technologies. Conventional methods for preparing desired two-photon polarization states are realized through bulky nonlinear crystals, which can restrict the versatility and tunability of the generated quantum states due to the fixed crystal nonlinear susceptibility. Here we present a solution using a nonlinear metasurface incorporating multiplexed silica metagratings on a lithium niobate film of 300 nanometer thickness. We fabricate two orthogonal metagratings on a single substrate with an identical resonant wavelength, thereby enabling the spectral indistinguishability of the emitted photons, and demonstrate in experiments that the two-photon polarization states can be shaped by the metagrating orientation. Leveraging this essential property, we formulate a theoretical approach for generating arbitrary polarization-entangled qutrit states by combining three metagratings on a single metasurface, allowing the encoding of desired quantum states or information. Our findings enable miniaturized optically controlled quantum devices using ultrathin metasurfaces as polarization-entangled photon sources.
\end{abstract}


The polarization of photons is an essential degree of freedom for understanding complex quantum phenomena and preparing desired quantum states for various quantum technologies. One can encode quantum information with the superposition of two orthogonal polarization states of a photon or produce polarization entanglement with multiple photons~\cite{Flamini:2019-16001:RPP}. These features of photon polarization enable the investigation of fundamental physics as well as state-of-the-art quantum applications including simulations~\cite{Georgescu:2014-153:RMP}, computations~\cite{Walther:2005-169:NAT, Prevedel:2007-65:NAT, Zhong:2020-1460:SCI}, and communications~\cite{Kim:2001-1370:PRL, Chou:2007-1316:SCI, Chen:2021-214:NAT, Ursin:2007-481:NPHYS, Liao:2017-43:NAT}. For example, polarization entanglement has been employed to demonstrate 
delayed-choice experiments~\cite{Ma:2012-479:NPHYS}, quantum key distribution~\cite{Wengerowsky:2018-225:NAT}, Bosonic-Fermionic quantum walk~\cite{Sansoni:2012-10502:PRL}, and to achieve high orbital angular momentum entanglement~\cite{Fickler:2012-640:SCI}.  In particular, higher dimensional quantum systems such as polarization qutrit, a three-level bosonic system, offer unique advantages over a qubit system~\cite{Lanyon:2008-60504:PRL}, allowing higher security in quantum information processing~\cite{Langford:2004-53601:PRL, Fujiwara:2003-167906:PRL}, more efficient quantum gates~\cite{Ralph:2007-22313:PRA}, and facilitating fundamental tests of quantum mechanics~\cite{Thew:2004-10503:PRL, Collins:2002-40404:PRL}. 

For free-space applications, generating polarization entanglement is conventionally realized with nonlinear crystals~\cite{Krivitskii:2005-521:JETP, Li:2015-28792:OE, Chen:2018-200502:PRL, Bogdanov:2004-230503:PRL, Vallone:2007-12319:PRA, Lanyon:2008-60504:PRL, Chekhova:2021:PolarizationLight} through the spontaneous parametric down-conversion (SPDC), where a pump photon splits into two photons called signal and idler.
This approach, however, utilizes millimeter- or centimeter-scale bulky crystals and manifests two limitations: (i)~dedicated temperature control is required to satisfy the momentum conservation of photons~\cite{Vallone:2007-12319:PRA, Lanyon:2008-60504:PRL} and (ii)~the polarization of generated photon pairs is intrinsically defined by the nonlinear tensor of the material~\cite{Okoth:2019-263602:PRL, Sultanov:2022-3872:OL}. As a result, the quantum states generated from the bulky crystals present limited versatility and tunability. Whereas recent work circumvented the limitation (i) by preparing the polarization entanglement with a thin nonlinear film~\cite{Sultanov:2022-3872:OL}, it came at the expense of a strongly reduced generation rate compared to bulk crystals and the constraint (ii) remained outstanding.

The recent developments of nonlinear dielectric metasurfaces, nanofabricated structures of sub-wavelength thickness incorporating nonlinear materials~\cite{Huang:2020-126101:RPP, DeAngelis:2020:NonlinearMetaOptics, Vabishchevich:2023-B50:PRJ, Neshev:2023-26:NPHOT}, show remarkable potential in realizing highly efficient and flexible miniaturized quantum light sources~\cite{Wang:2022-38:PT, Kan:2023-2202759:ADOM}. 
As metasurfaces can dramatically enhance and tailor nonlinear light-matter interactions through the optical resonances supported by the nanostructures~\cite{Fedotova:2020-8608:NANL, Yuan:2021-153901:PRL}, they have been demonstrated for strong enhancement of photon-pair generation through SPDC~\cite{Santiago-Cruz:2021-4423:NANL, Zhang:2022-eabq4240:SCA, Santiago-Cruz:2022-991:SCI, Son:2023-2567:NASC} and the preparation of complex quantum states featuring spectral~\cite{Santiago-Cruz:2022-991:SCI} or spatial~\cite{Zhang:2022-eabq4240:SCA, zhang2023photon} entanglement. 
However, the controllable generation of biphoton polarization states with metasurfaces was unexplored.

In this work, we present a general approach for engineering polarization states of photon pairs generated from a nonlinear metasurface and preparing arbitrary polarization Bell states and qutrits optically reconfigurable via the pump laser beam. We experimentally develop a single metasurface incorporating multiplexed metagratings on top of an \textit{x}-cut LiNbO$_3$ thin film which features strong quadratic nonlinearity. All metagratings can resonantly enhance the generation of photon pairs linearly polarized along the distinct groove directions, thereby overcoming the fundamental limitation of unpatterned LiNbO$_3$ crystals, as experimentally confirmed by the full polarization state tomography. We show theoretically that pumping a metasurface containing three metagratings with horizontal, vertical, and diagonal orientations by a spatially modulated coherent laser allows the tunable generation of arbitrary polarization entanglement. These results open the door to the miniaturization of various optically reconfigurable quantum devices based on ultra-thin metasurfaces as room-temperature polarization-entangled photon sources.

\begin{figure*}[ht]   
\centering
\fbox{\includegraphics[width=0.81\linewidth]{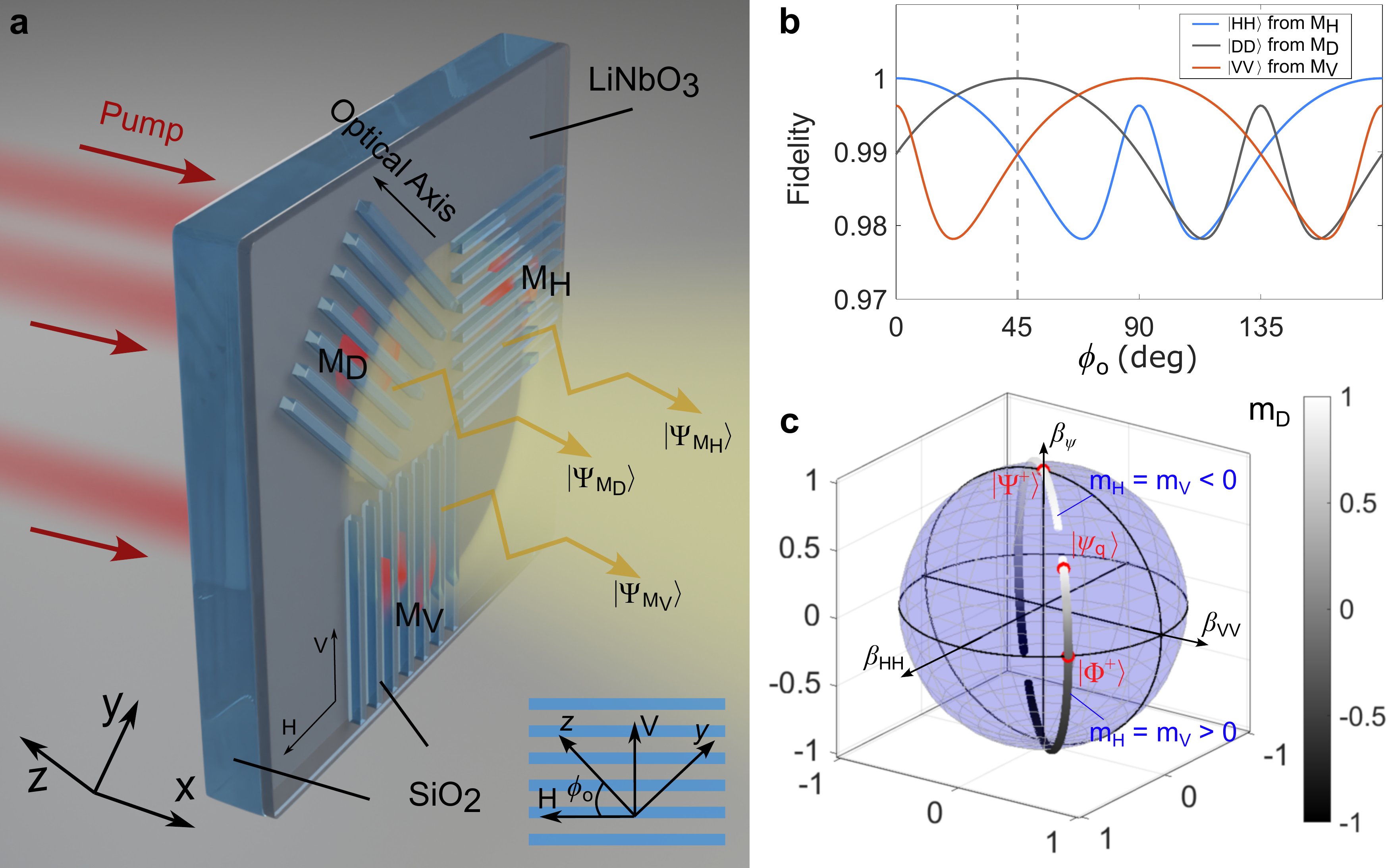}}
\caption{\textbf{Concept.} \textbf{a} Sketch of generation of polarization-entangled photon pairs with multiplexed metagratings in a metasurface. A LiNbO$_3$ film of 300~nm thickness is covered by a metagrating ($\rm{M_{\rm D}}$) oriented along its optical axis ($z-$axis) and two orthogonal metagratings ($\rm{M_{\rm H}}$, $\rm{M_{\rm V}}$) placed along horizontal and vertical directions respectively (45$^{\circ}$ with respect to the $z-$axis). \textbf{b} Fidelity of polarization states generated from metagratings ($M_{\rm H}$, $M_{\rm D}$ and $M_{\rm V}$) vs. optical axis orientation angle $\phi_{\rm o}$. The fidelity is larger than 97\% for any angle $\phi_{\rm o}$, indicating that the photon polarization is determined by the metagrating orientation. The enhancement factor is set as $\zeta = 10^4$. \textbf{c} Coefficients $\beta_{\rm HH}$, $\beta_{\rm VV}$, and $\beta_{\psi}$ of polarization state $\ket{\Psi_{\rm M}}$ produced from the metasurface vs. coefficient $m_{\rm D}$ of photon state $\ket{\Psi_{{\rm M_{\rm D}}}}$ from metagrating ${\rm M_{\rm D}}$. Here we set the coefficients $m_{\rm H} = m_{\rm V} = \pm\sqrt{(1-m_{\rm D}^2)/2}$. The red dots indicate three different maximally-entangled states, i.e., $\ket{\Phi^+}=(\ket{H_sH_i}+\ket{V_sV_i})/\sqrt{2}$ at $m_{\rm H}=m_{\rm V}=0.69$, $\ket{\Psi^{+}} =\left(\ket{H_{{\rm s}}V_{{\rm i}}} +\ket{V_{{\rm s}}H_{{\rm i}}} \right) /\sqrt{2}$ at $m_{\rm H}=m_{\rm V}=-0.41$, and $\ket{\psi_q}=(\ket{H_sH_i}+\ket{V_sV_i}+\ket{\Psi^{+}}) /\sqrt{3}$ at $m_{\rm H}=m_{\rm V}=0.24$}.
\label{fig:sketch}
\end{figure*}

Metasurfaces based on the LiNbO$_3$-on-insulator platform were recently developed for various applications including electro-optic modulation, classical optical frequency conversion, and photon-pair generation~\cite{Fedotova:2022-3745:ACSP}. With this platform, we engineer and multiplex polarization state of photon pairs generated from a metasurface where periodic SiO$_2$ metagratings are fabricated on top of a subwavelength-thin layer of  LiNbO$_3$. Specifically, we consider a single metasurface incorporating three linear metagratings ($\rm{M_{\rm H}, M_{\rm V}, M_{\rm D}}$) with their orientations along the horizontal, vertical, and diagonal directions, respectively, as shown in Fig.~\ref{fig:sketch}\textbf{a}. We choose an $x$-cut LiNbO$_3$ film where its optical ($z$-) axis is oriented along the direction of the diagonal metagrating ($\rm{M_{\rm D}}$) and by 45~degrees with respect to the directions of both horizontal ($\rm{M_{\rm H}}$) and vertical ($\rm{M_{\rm V}}$) metagratings. 

To depict the physics of polarization engineering of photon pair generation enabled by SPDC from the metasurface, we first analyse the classical inverse nonlinear process of sum-frequency generation (SFG) in the LiNbO$_3$ film. In SFG, the $z$-component of nonlinear polarization generated from fundamental beams (signal/idler photons) propagating along the $x$-axis is determined as $P^{NL}_z = 4\epsilon_0(d_{33}E_z^sE_z^i + d_{31}E_y^sE_y^i)$, where $\epsilon_0$ is the vacuum permittivity and $E_{y,z}^{s,i}$ are the electric fields of fundamental beams. The nonlinear tensor coefficients are $d_{33} = 19.5$ pm/V and $d_{31} = 3.2$  pm/V around the 1313~nm wavelength~\cite{Shoji:1997-2268:JOSB}. According to the classical-to-quantum correspondence~\cite{Poddubny:2020-147:QuantumNonlinear}, the thin film will generate photon pairs at the following polarization state via the SPDC process when a $z$-polarized pump laser is incident on the film, 
$\ket{\Psi_{{\rm film}}} =\left({d_{33}^{2}+d_{31}^{2}}\right)^{-1/2} \left(d_{33}\ket{Z_{{\rm s}}} \otimes\ket{Z_{{\rm i}}} +d_{31}\ket{Y_{{\rm s}}} \otimes\ket{Y_{{\rm i}}} \right),$
where $\ket{Z_{s,i}}$ and $\ket{Y_{s,i}}$ indicate the polarization states of signal and idler photons. Since $d_{33} \gg d_{31}$, the film predominantly emits photon pairs with a fixed linear polarization along $z$-axis, which is a fundamental limitation of unstructured LiNbO$_3$ crystals. 

Here we overcome the material limitation by engineering the orientation of gratings that support guided-mode resonances with their excitation dependent on the polarization of signal and idler photons. We initially focus on the metagrating $M_{\rm H}$ where the grating orientation is defined as ``Horizontal'' ($H$) and the angle of the optical $z$-axis with respect to the $H$-direction is denoted as $\phi_{\rm o}$, as illustrated in the inset of Fig.~\ref{fig:sketch}\textbf{a}. The polarization states $\ket{Z_{s,i}}$ and $\ket{Y_{s,i}}$ of signal and idler photons from the unstructured film can be 
expressed in the rotated coordinate system aligned with the grating as
$\ket{Z_{s,i}} = \cos(\phi_{\rm o})\ket{H_{s,i}} +\sin(\phi_{\rm o})\ket{V_{s,i}}$ and $\ket{Y_{s,i}} = -\sin(\phi_{\rm o})\ket{H_{s,i}} +\cos(\phi_{\rm o})\ket{V_{s,i}}$.
The optical resonances supported by metagratings selectively enhance the photon-pair generation along the direction of the grooves. Specifically, the horizontal components $\ket{H_{s,i}}$ would be effectively multiplied by a factor $\zeta^{1/4}$, where $\zeta$ denotes the enhancement of photon pair rate from the metasurface compared to an unpatterned film in the case that the grating is oriented along the optical axis. As a result, the normalized polarization state of photon pairs emitted from the metasurface is expressed as (see Supporting Information Sec.~S1 for details)
\begin{equation} \label{eq:metaH}
\ket{\Psi_{{\rm M_{\rm H}}}} =  \frac{\alpha_{{\rm HH}}\ket{H_{{\rm s}}H_{{\rm i}}} +\alpha_{{\rm HV}}\ket{H_{{\rm s}}V_{{\rm i}}} +\alpha_{{\rm VH}}\ket{V_{{\rm s}}H_{{\rm i}}} +\alpha_{{\rm VV}}\ket{V_{{\rm s}}V_{{\rm i}}}}{\sqrt{|\alpha_{{\rm HH}}|^{2}+|\alpha_{{\rm HV}}|^{2}+|\alpha_{{\rm VH}}|^{2}+|\alpha_{{\rm VV}}|^{2}}} ,
\end{equation}
where $\alpha_{{\rm HH}} =\sqrt{\zeta}[(d_{33}\cos^{2}(\phi_{\rm o})+d_{31}\sin^{2}(\phi_{\rm o})]$, $\alpha_{{\rm HV}} = \alpha_{{\rm VH}} =\zeta^{1/4}(d_{33}-d_{31})\cos(\phi_{\rm o})\sin(\phi_{\rm o})$, 
$\alpha_{{\rm VV}} = d_{33}\sin^{2}(\phi_{\rm o})+d_{31}\cos^{2}(\phi_{\rm o})$. Notably, the SPDC processes associated with $d_{31}$ that are conventionally neglected in previous works~\cite{Zhang:2022-eabq4240:SCA, Santiago-Cruz:2021-4423:NANL, Santiago-Cruz:2022-991:SCI, Parry:2021-55001:ADP} due to $d_{33} \gg d_{31}$ play an indispensable role in engineering the polarization states, such that the photon polarization is always closely parallel to the metagrating grooves. Specifically, as a result of the combined contribution of $d_{31}$ and $d_{33}$, the pairs are in a polarization state $\ket{H_{s}H_{i}}$ with a fidelity higher than 97\% for an arbitrary angle $\phi_{\rm o}$ as shown in Fig.~\ref{fig:sketch}\textbf{b}. Here an enhancement factor $\zeta=10^4$ is considered, based on the ratio $\sqrt{\zeta}=\alpha_{\rm HH}/\alpha_{\rm VV}$ obtained via degenerate SFG simulation on metasurface $M_{\rm H}$ with $\phi_{\rm o}=45^\circ$.
Accordingly, we can produce photons with desired polarization states by engineering the grating orientation.

We now analyse the configuration of three metagratings as sketched in Fig.~\ref{fig:sketch}{\bf a}. Each grating can be pumped individually by splitting a laser beam into three coherent spatial modes with independently controlled amplitudes and phases, which can be realized with a spatial light modulator.
Considering frequency-degenerate SPDC where the spatial and frequency spectra of the photon pairs from the metagratings are identical, the polarization state from the metasurface is the superposition of quantum states produced from each grating, i.e., $\ket{\Psi_{{\rm M}}} =m_{\rm H}\ket{\Psi_{\rm M_{\rm H}}} +m_{\rm V}\ket{\Psi_{{\rm M_{\rm V}}}}+m_{\rm D}\ket{\Psi_{{\rm M_{\rm D}}}}$, where $m_{\rm  H, V, D}$ are the relative complex amplitudes of pump beams incident on three metagratings satisfying the normalization condition $|m_{\rm H}|^2 + |m_{\rm V}|^2 + |m_{\rm D}|^2 = 1$. Explicit expressions for the states $\ket{\Psi_{{\rm M_{\rm V}}}}$ and $\ket{\Psi_{{\rm M_{\rm D}}}}$ from vertical ($M_{\rm V}$) and diagonal ($M_{\rm D}$) metagratings are derived using the same approach as for $\ket{\Psi_{{\rm M_{\rm H}}}}$, and provided in the Supporting Information Sec.~S2. The state $\ket{\Psi_{{\rm M}}}$ can be represented in terms of conventional polarization qutrit basis,

\begin{equation} \label{eq:metaM}
\ket{\Psi_{{\rm M}}} = \mathcal{N} \left(\beta_{{\rm HH}}\ket{H_{{\rm s}}H_{{\rm i}}} +\beta_{{\rm VV}}\ket{V_{{\rm s}}V_{{\rm i}}} +\beta_{\psi}\ket{\Psi^{+}} \right) ,
\end{equation}
where $\beta_{{\rm HH,VV},\psi}$ denote the complex coefficients of each basis and $\mathcal{N} = 1/\sqrt{|\beta_{{\rm HH}}|^{2}+|\beta_{{\rm VV}}|^{2}+|\beta_{{\psi}}|^{2}}$ is the normalization factor. The last term $\ket{\Psi^{+}} ={1}/{\sqrt{2}}\left(\ket{H_{{\rm s}}V_{{\rm i}}} +\ket{V_{{\rm s}}H_{{\rm i}}} \right)$ is one of the polarization Bell states, which is orthogonal to $\ket{H_{{\rm s}}H_{{\rm i}}}$ and $\ket{V_{{\rm s}}V_{{\rm i}}}$. 

Our key finding based on the preceding analysis is that it is possible to produce any desired polarization qutrit by tuning the pump amplitudes and phases as $[m_{\rm H},m_{\rm V}, m_{\rm D}]^T = \mathcal{M}^{-1}[\beta_{HH}, \beta_{HV}, \beta_{\Psi}]^T$, where the matrix $\mathcal{M}$ is defined in the Supporting Information Sec.~S2. For illustration, we consider two special cases in Fig.~\ref{fig:sketch}\textbf{c}, where we set $m_{\rm H}=m_{\rm V}=\pm\sqrt{(1-m_{\rm D}^2)/2}$ and vary the coefficient $m_{\rm D}$ such that  entanglement of interest can be produced.
We present the results on a sphere surface where $|\beta_{{\rm HH}}|^{2}+|\beta_{{\rm VV}}|^{2}+|\beta_{{\psi}}|^{2}=1$. At specific combinations of pumping coefficients, one can produce maximally entangled polarization states, such as Bell states $\ket{\Phi^+}=\frac{1}{\sqrt{2}}(\ket{H_sH_i}+\ket{V_sV_i})$ or $\ket{\Psi^{+}}$ and a qutrit $\ket{\psi_q}=\frac{1}{\sqrt{3}}(\ket{H_sH_i}+\ket{V_sV_i}+\ket{\Psi^{+}})$, as indicated by red dots in Fig.~\ref{fig:sketch}\textbf{c}.

Next, we focus on the experimental fabrication and characterization of the metasurface. We employ metagratings incorporating nonlocal guided-wave resonances~(see Supporting Information Sec.~S3), which support high-Q modes facilitating the generation of near-degenerate photon pairs with narrow  spectra across a range of pump wavelengths~\cite{Zhang:2022-eabq4240:SCA}. Previously, we demonstrated the generation of photon pairs from a metagrating orientated along the optical axis of LiNbO$_3$ thin film~\cite{Zhang:2022-eabq4240:SCA}, corresponding to $M_{\rm D}$ in Fig.~\ref{fig:sketch}{\bf a}. In that case, the photons were emitted with the same linear polarization along the optical axis both from an unstructured thin film and the metagrating. Here, we experimentally demonstrate that the polarization state of photon pairs can be  nontrivially engineered by the metagrating orientations that are significantly different from the optical axis. Specifically, we consider a metasurface incorporating two crossed linear metagratings oriented by $45^{\circ}$ and $-45^{\circ}$ with respect to the optical axis, i.e. $M_{\rm H}$ and $M_{\rm V}$. Whereas three gratings are needed to generate arbitrary qutrits as discussed above, a combination of these two gratings can already be used to generate a range of polarization states, including the maximally entangled one $\ket{\Phi^+}=(\ket{H_sH_i}+\ket{V_sV_i})/\sqrt{2}$, which is not possible from an unstructured lithium niobate thin film.

\begin{figure*}[ht]
\centering
\fbox{\includegraphics[width=0.48\linewidth]{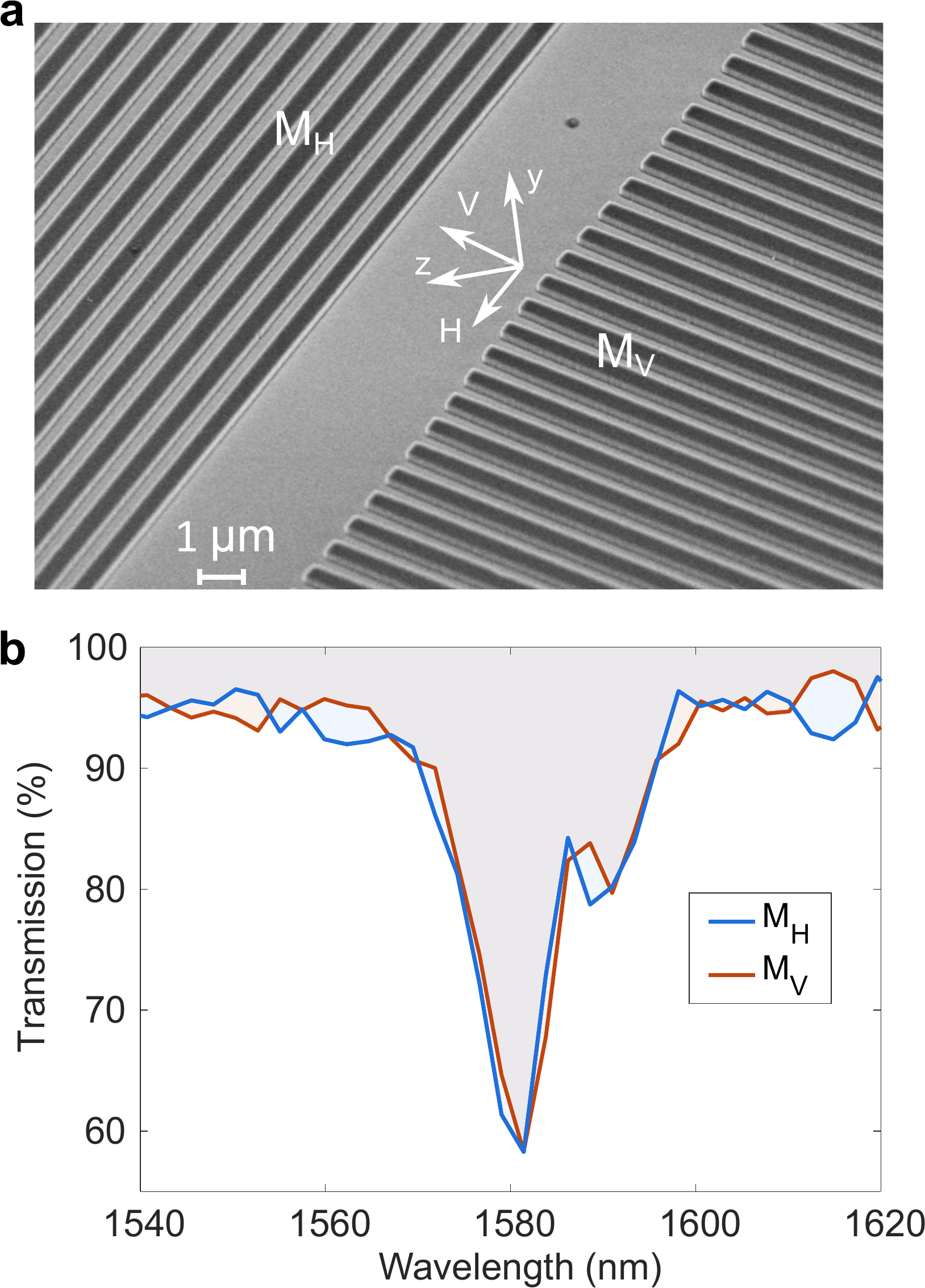}}
\caption{\textbf{Fabrication and classical characterization of metasurface.} \textbf{a} SEM image of fabricated dielectric meta-gratings with directions orthogonal to each other. \textbf{b} Measured linear transmission as a function of wavelength at normal incidence, for the input polarization along the corresponding grating directions. Both horizontal ($M_{\rm H}$) and vertical ($M_{\rm V}$) metagratings have very similar resonance shapes around 1581~nm.
}
\label{fig:SEM}
\end{figure*}

Due to the identical absolute angles of $|\pm 45^{\circ}|$ for the metagratings $M_{\rm H}$ and $M_{\rm V}$ with respect to the crystal axis, theoretically their linear spectra should be identical if the gratings are the same. Accordingly, we have designed both gratings with the same period and shape of the unit cell to support a resonance at normal incidence around the target photon generation wavelengths of $1580$~nm. Then, we performed the fabrication with electron beam lithography and etching processes (Supporting Information Sec.~S4).

The scanning electron microscopy (SEM) image of the fabricated metasurface is shown in Fig.~\ref{fig:SEM}\textbf{a}. The dimensions of two gratings are the same, with a thickness of 215~nm, period of  884~nm, and filling ratio of~66\%. 
We experimentally measure the normal-incident transmission for both metagratings using a spectrometer with resolution $\sim$2.4~nm, shown in Fig.~\ref{fig:SEM}\textbf{b}, pumped with a broadband tungsten-halogen lamp source polarized along the corresponding grating directions. The optical resonances supported by the two metasurfaces manifest at 1581~nm. The close similarity of the transmission spectra can facilitate the generation of indistinguishable photon pairs from the metasurfaces in the frequency domain. 

\begin{figure*}[!t]
\centering
\fbox{\includegraphics[width=0.45\linewidth]{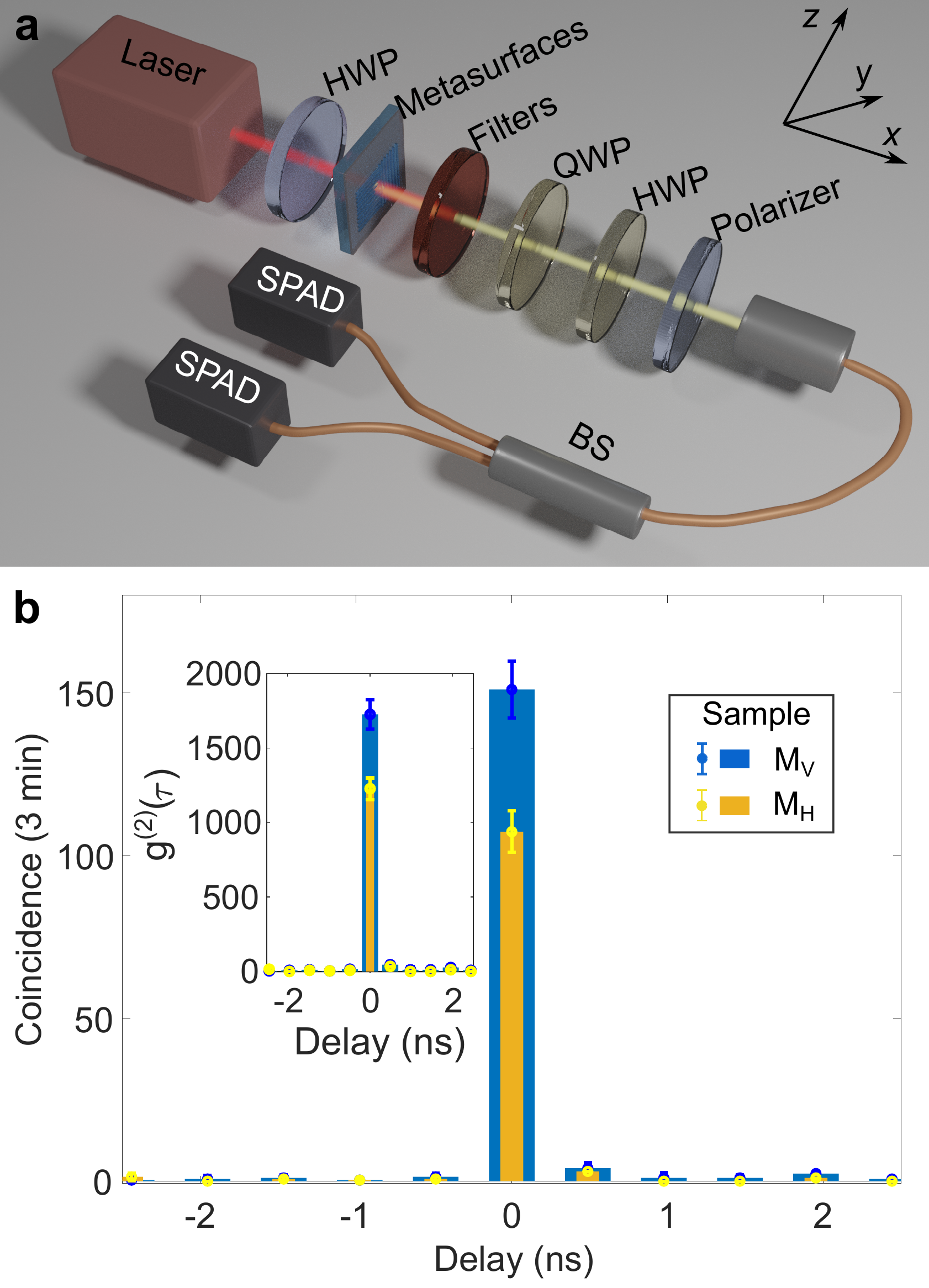}}
\caption{\textbf{Quantum characterization of correlated photon pairs.} \textbf{a} Experimental setup. A continuous-wave laser is focused on the nonlinear metasurface, generating photon pairs with desired polarization states. Photon pairs passing through a 50:50 fiber beam splitter (BS) are detected by two single photon detectors (SPAD), where the two-photon correlation is analyzed with the coincidental arrival events of a pair of photons. Polarization states of photon pairs are characterized by a quarter-wave plate (QWP), a half-wave plate (HWP), and a polarizer oriented along $H$-direction. \textbf{b} Coincidence measurement of photon pairs from metagratings $M_{\rm H}$ and $M_{\rm V}$. The inset shows the second correlation function $g^{(2)}(\tau)$ as a function of two-photon delays. The QWP angle is 0 for both metagratings and the HWP angle is 0 for $M_{\rm H}$ and 45$^o$ for $M_{\rm V}$.}
\label{fig:coin}
\end{figure*}

The quantum photon pairs generated from $M_{\rm H}$ and $M_{\rm V}$ through SPDC are characterized by the setup sketched in Fig.~\ref{fig:coin}\textbf{a}, with detailed methods given in Supporting Information Sec.~S4.
We present the measured coincidence counts of photon pairs from the metasurfaces for various time delays between two detectors in Fig.~\ref{fig:coin}\textbf{b}. At a pump power of 85~mW, the counts over an integration time of 3~min are 110 and 150 for metasurfaces $\rm{M_{\rm H}}$ and $\rm{M_{\rm V}}$, corresponding to a rate of 0.61~Hz and 0.83~Hz, respectively. We attribute the difference in SPDC efficiency to slight misalignment during the fabrication process, whereby the angle $\phi_{\rm o}$ between the grating orientation and the optical axis is not exactly $45^{\circ}$. In terms of $\alpha_{\rm HH}$ in Eq.~(\ref{eq:metaH}), the rate ratio of photon pairs from respective metasurfaces is 
\begin{equation} \label{eq:Ratio_VH}
    R_{V/H} =\left|\frac{d_{33}\tan^{2}(\phi_{\rm o})+d_{31}}{d_{33}+d_{31}\tan^{2}(\phi_{\rm o})}\right|^{2} .
\end{equation}
We substitute the measured ratio of $R_{V/H} = 150/110$ in Eq.~(\ref{eq:Ratio_VH}) to estimate the angle as $\phi_{\rm o} \simeq 48^{\circ}$, which is only $\sim 3^{\circ}$ off from the theoretical design. The difference in rates 
can be simply compensated by changing the pump laser powers incident on individual gratings. 
For comparison, we measure the maximum photon pair rate from an unpatterned film as $4$~mHz under the same experimental conditions, indicating that the SPDC efficiency is enhanced by 150 and 210 times in metasurfaces $M_{\rm H}$ and $M_{\rm V}$, respectively.

\begin{figure*}[ht]
\centering
\fbox{\includegraphics[width=0.98\linewidth]{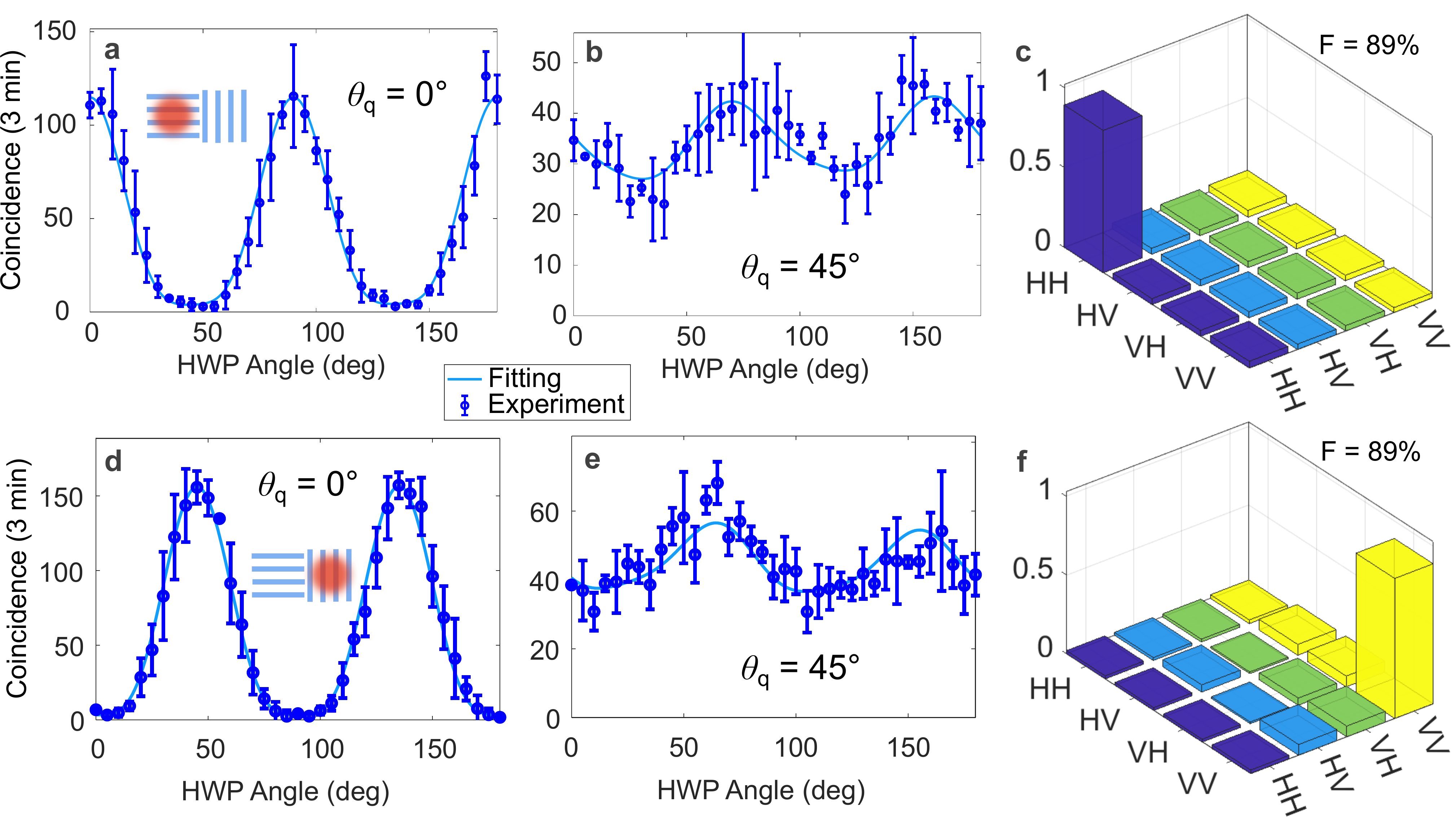}}
\caption{\textbf{Experimental demonstration: polarization engineering by metagrating orientation.} Quantum polarization state tomography for the pump illuminating metagrating \textbf{a-c} $\rm{M_{\rm H}}$ or \textbf{d-f} $\rm{M_{\rm V}}$ on the same metasurface.
\textbf{a-b,d-e} Coincidence measurement as a function of HWP angle for different QWP angles $\theta_q$ as indicated by labels.
\textbf{a,d}~The visibility is
higher than $90\%$ at $\theta_q = 0^{\circ}$, while in \textbf{b,e} weak modulations are observed at $\theta_q = 45^{\circ}$. \textbf{c,f}~The density matrix is reconstructed from the correlation measurements in the respective figure rows, indicating a polarization state of \textbf{c}~$\ket{HH}$ with a fidelity of $\sim 89\%$ or \textbf{f}~$\ket{VV}$ with a fidelity of $\sim 89\%$.
}
\label{fig:Tomo}
\end{figure*}


We perform the quantum tomography of polarization states of photon pairs~\cite{James:2001-52312:PRA} generated from individual metagratings by characterizing coincidence rates at different combinations of half-wave plate (HWP) and quarter-wave plate (QWP) angles while fixing the polarizer orientation to the H-direction. 
Figures \ref{fig:Tomo}\textbf{a}-\textbf{b} and \ref{fig:Tomo}\textbf{d}-\textbf{e} present the coincidence measurement for a varying HWP angle $\theta_h$ at two QWP angles $\theta_q$. In the case of $\theta_q = 0$, H or V linearly polarized photons would remain unchanged after traveling through the QWP and then would be partially filtered out by the polarizer depending on the HWP angle. By comparing Figs.~\ref{fig:Tomo}\textbf{a} and~\textbf{d}, it is found that the polarization states of photon pairs from two crossed metasurfaces are orthogonal as expected. Note that these dependencies are distinct from the case of classical light or a single photon when it would have the cosine squared shape. When the QWP angle is set at $\theta_q = 45^\circ$, purely H or V polarized photon pairs would turn into a circularly polarized beam after the QWP, and then the coincidence rate should be independent on the HWP angle. The weak modulations in Figs.~\ref{fig:Tomo}\textbf{b} and~\textbf{e} imply that generated photons may involve smaller cross-polarized 
components in addition to a $\ket{HH}$ or $\ket{VV}$ state. 

We now perform the reconstruction of the density matrix $\rho$ of the two-photon polarization states. Photon pairs produced from metasurfaces travel through a QWP and a HWP, after which its polarization state is projected to the horizontal direction. The overall projection state is thus given as $\ket{\Psi_{{\rm proj}}} =\hat{U}_{{\rm QWP}}\hat{U}_{{\rm HWP}}\ket{H}$, where $\hat{U}_{{\rm QWP}}$, $\hat{U}_{{\rm HWP}}$ and $\ket{H}$ can be expressed using Jones Matrices (Supporting Information Sec.~S5). As a result, the expected photon pair rate is
\begin{equation} \label{eq:RateFit}
\varPi_{\rm exp} = {\rm Trace}\left(\rho\ket{\Psi_{\rm proj}} \bra{\Psi_{\rm proj}}\right),
\end{equation}
where $\rho$ is the qutrit density matrix that is defined by 10 free real-valued parameters~\cite{Titchener:2018-19:NPJQI}.
Using the nonlinear regression method, and enforcing the non-negative definiteness of the density matrix~\cite{James:2001-52312:PRA}, we fit the 10 parameters using Eq.~(\ref{eq:RateFit}) with the measured photon-pair rates of respective metagratings at all HWP and QWP angles. The blue solid curves in Figs.~\ref{fig:Tomo}\textbf{a}-\textbf{b} and~\textbf{d}-\textbf{e} are the fitting results 
for the metagratings $M_{\rm H}$ and $M_{\rm V}$, showing a consistent agreement between experiment and theory. The reconstructed density matrices of two-photon states from two crossed metagratings are shown in Figs.~\ref{fig:Tomo}\textbf{c} and~\textbf{f}. The fidelity $F = \left({\rm Trace}\left(\sqrt{\rho\ket{\psi_{\rm exp}}\bra{\psi_{\rm exp}}}\right)\right)^2$ with the expected state $\ket{\psi_{\rm exp}}$ ($\ket{\psi_{\rm exp}} = \ket{HH}$ for $M_{\rm H}$ and $\ket{\psi_{\rm exp}} = \ket{VV}$ for $M_{\rm V}$) are both 89\%. Because of spectral and spatial broadening in the experiment, the calibrated fidelity is smaller than the theoretically predicted values of $\sim$99\% at ideal conditions as shown in Fig.~\ref{fig:sketch}\textbf{b}.
We note that the photon pairs generated by metagrating $M_{\rm D}$ exhibit polarization along the groove direction with an extinction ratio exceeding 99\%, as demonstrated in our previous work~\cite{Zhang:2022-eabq4240:SCA}. The high fidelity obtained indicates that our nonlinear metasurface based on LiNbO$_3$ is a promising platform for manipulating polarization states of photon pairs and realizing polarization multiplexing.

\begin{figure*}[ht]
\centering
\fbox{\includegraphics[width=0.95\linewidth]{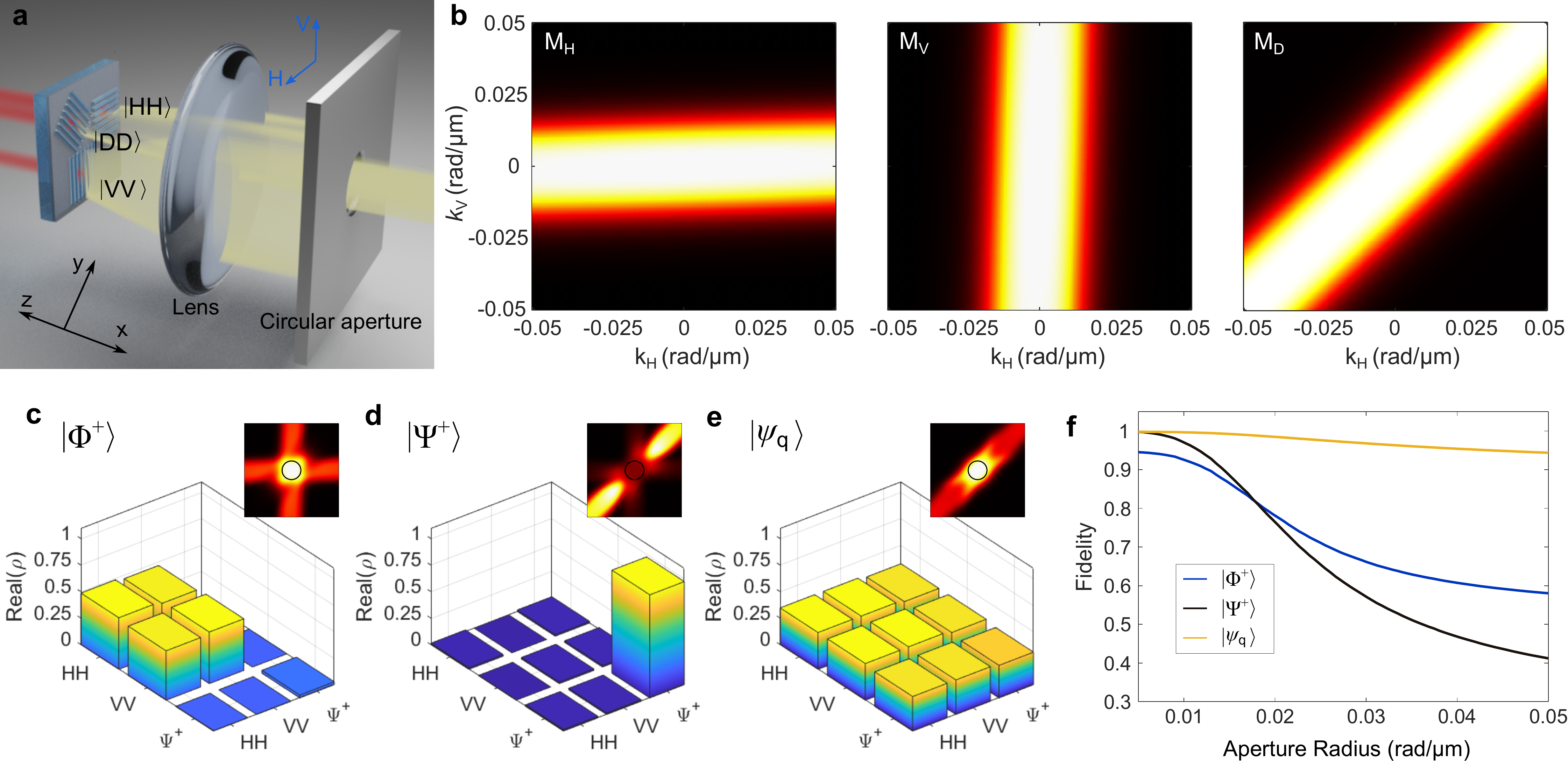}}
\caption{\textbf{Theoretical simulation: generation of polarization entanglement.} \textbf{a} Sketch of preparation of polarization entanglement. A circular aperture is added in the $k$-space to guarantee a high spatial overlap of photon emissions from three metagratings. \textbf{b} Emission pattern of each metagrating in $k$-space. The pump wavelength is set as 790.4~nm, which is half of the resonant wavelength of the bright mode at normal incidence. \textbf{c-e} Density matrices of quantum states from the metasurface. Three quantum states $\ket{\Phi^+}$, $\ket{\Psi^+}$ and $\ket{\psi_q}$ with high-degree of entanglement are produced at three sets of coefficients $m_{\rm H,V,D}$. The insets present the corresponding emission pattern and the black circles represent the spatial filtering implemented. The aperture radius is set as 0.01 rad/$\mu$m. The coefficients $m_{\rm H,V,D}$ used here are the same as in Figs.~\ref{fig:sketch}\textbf{b}-\textbf{c}. \textbf{f} Fidelity of quantum states vs. the aperture radius, showing a decrease at larger apertures due to reduced spatial overlaps.}
\label{fig:entangle}
\end{figure*}

Next, we discuss the theoretical approach for generating polarization entanglement from the metasurface, based on the simulation of far-field emission wavefunctions of photon pairs. We consider the configuration sketched in Fig.~\ref{fig:entangle}\textbf{a} where three metagratings (${\rm M_{\rm H}}$, ${\rm M_{\rm V}}$ and ${\rm M_{\rm D}}$) are pumped simultaneously, followed by a lens to transfer the emission pattern into $k$-space where a circular aperture is placed to select  spatially overlapping photons emitted from three metagratings. The wavefunction of photon pairs from each metagrating is calculated individually according to its reverse process~-- SFG via coupled mode theory~\cite{Suh:2004-1511:IQE, Sun:2022-1575:PRJ, Zhang:2022-eabq4240:SCA} and quantum-classical correspondence~\cite{Marino:2019-1416:OPT, Parry:2021-55001:ADP} (Supporting Information Sec.~S6). The emission pattern of each metagrating is defined by its orientation, as displayed in Fig.~\ref{fig:entangle}\textbf{b}. The patterns are almost parallel to the grooves, with those from ${\rm M_{\rm H}}$ and ${\rm M_{\rm V}}$ being slightly tilted due to material birefringence. Here we consider a pump frequency equal to twice the resonance at normal incidence to facilitate the optimal spatial overlap between emissions from all three metagratings. 

According to our concept, different Bell states (e.g., $\ket{\Phi^+}$ and $\ket{\Psi^+}$) and maximally-entangled qutrit $\ket{\psi_q}$ may be produced with particular combinations of coefficients $m_{\rm H, V, D}$, as we have illustrated in Figs.~\ref{fig:sketch}\textbf{c-e}. Appropriate spatial filtering (black circle) is added to the total emission (insets) to achieve a higher spatial overlap. The predicted density matrices are shown in Figs.~\ref{fig:entangle}\textbf{c-e}, which closely match the target polarization-entangled states.
Additionally, we calculate the effect of the aperture on the quantum fidelity, which indicates the closeness between the produced and target quantum states, as shown in Fig.~\ref{fig:entangle}\textbf{f}. We find that the fidelity of qutrit $\ket{\psi_q}$ remains above~94\% for a reasonably large aperture radius while the high fidelity of $\ket{\Psi^+}$ demands a smaller aperture. This is because the spatial emission overlap of each basis $\ket{H_sH_i}$, $\ket{V_sV_i}$ and $\ket{\Psi^+}$ is higher for state $\ket{\psi_q}$ but lower for $\ket{\Psi^+}$, as indicated by the emission patterns of each quantum states shown in the insets of Figs.~\ref{fig:entangle}\textbf{c-e} (Supporting Information Sec.~S6). 

In summary, we have developed a general approach for manipulating polarization states of emitted photon pairs by engineering the orientation of the metagratings on a nonlinear metasurface, beyond the natural constraints of unpatterned nonlinear crystals. The production of photon pairs with different polarizations from the lithium niobate metasurface is experimentally confirmed with quantum state tomography. By employing the developed technique of polarization engineering, we propose a theoretical approach to generate optically controllable qutrits. Specifically, we show that the designed metasurface containing multiple metagratings with different orientations can generate a high-fidelity qutrit with an arbitrary density matrix that is optically tunable via the pump beams shaped by a spatial light modulator. Such miniaturized metasurface photon-pair sources
provide a novel approach to optically encode and engineer quantum information, paving the way for numerous quantum applications 
spanning quantum communications, quantum information processing, and quantum computing. 

\begin{suppinfo}
The Supporting Information is available free of charge at

Extra details on polarization engineering of photon pairs, generation of polarization qutrit, guided-mode resonances for three metagratings, quantum tomography protocol, theoretical analysis of qutrit generation with far-field emission wavefunctions, and experimental methods.

\end{suppinfo}

\section{Author Contributions}
J.M. and J.Z. contributed equally.

\section{Note}
The authors declare no competing financial interest.

\begin{acknowledgement}
This work was supported by the Australian Research Council (DP190101559, CE200100010). The metasurface fabrication was performed at the Australian National University node of the Australian National Fabrication Facility (ANFF), a company established under the National Collaborative Research Infrastructure Strategy to provide nano and micro-fabrication facilities for Australia’s researchers.

\end{acknowledgement}


\pagebreak

\providecommand{\latin}[1]{#1}
\makeatletter
\providecommand{\doi}
  {\begingroup\let\do\@makeother\dospecials
  \catcode`\{=1 \catcode`\}=2 \doi@aux}
\providecommand{\doi@aux}[1]{\endgroup\texttt{#1}}
\makeatother
\providecommand*\mcitethebibliography{\thebibliography}
\csname @ifundefined\endcsname{endmcitethebibliography}
  {\let\endmcitethebibliography\endthebibliography}{}



\end{document}